\documentclass[useAMS,usenatbib]{mn2e}
\usepackage{graphicx}
\usepackage{amsmath}

\newcommand{\nitrogen}{[N~{\sc ii}]}
\newcommand{\nitrogena}{[N~{\sc i}]}
\newcommand{\oxygeniii}{[O~{\sc iii}]}
\newcommand{\oxygeni}{[O~{\sc i}]}
\newcommand{\oxygenii}{[O~{\sc ii}]}

\newcommand{\sulfurt}{[S~{\sc ii}]}

\newcommand{\ironii}{[Fe~{\sc ii}]}

\def\vhel{\ifmmode{V_{{\rm HEL}}}\else{$V_{{\rm HEL}}$}\fi}
\def\vsys{\ifmmode{V_{\rm sys}}\else{$V_{\rm sys}$}\fi}
\def\kms{\ifmmode{~{\rm km\,s}^{-1}}\else{~km~s$^{-1}$}\fi}
\def\vlsr{\ifmmode{v_{\rm lsr}}\else{$v_{\rm lsr}$}\fi}

\title[H$_2$ in low-ionization structures of PNe]
{H$_2$ in low-ionization structures of planetary nebulae \thanks{Based on observations obtained at the Gemini 
Observatory, which is operated by the Association of Universities for Research in Astronomy, 
Inc., under a cooperative agreement with the NSF on behalf of the Gemini partnership.}} 

\author[Akras, Gon\c calves \& Ramos-Larios] 
{Stavros Akras$^{1}$ $\thanks{e-mail:akras@astro.ufrj.br}$,
Denise R. Gon\c calves$^{1}$ $\thanks{e-mail:denise@astro.ufrj.br}$, Gerardo Ramos-Larios$^{2}$\\
$^{1}$ Observat\'{o}rio do Valongo, Universidade Federal do Rio de Janeiro, Ladeira Pedro Antonio 43, 20080-090, Rio de Janeiro, Brazil\\
$^{2}$ Instituto de Astronom\'{i}a y Meteorolog\'{i}a, Av. Vallarta No. 2602. Col. Arcos Vallarta, CP 44130, Guadalajara, Jalisco, Mexico}

\begin{document}  

\date{Received **insert**; Accepted **insert**}

\pagerange{\pageref{firstpage}--\pageref{lastpage}}

\maketitle
\label{firstpage}

\begin{abstract}

We report the detection of near-IR H$_2$ emission from the low-ionization structures (knots) in two planetary nebulae. The deepest ever high-angular resolution H$_2$ 1-0 S(1) at 2.122~\micron, H$_2$ 2-1 S(1) at 2.248~\micron\ and Br$\gamma$ images of K~4-47 and NGC~7662, obtained using the Near InfraRed Imager and Spectrometer (NIRI) at Gemini-North, are analyzed here. K~4-47 reveals a remarkable highly collimated bipolar structure not only 
in the optical but also in the molecular hydrogen emission. The H$_2$ emission emanates from the walls of the bipolar outflows and also from the pair of knots at the tip of the outflows. The H$_2$ 1-0 S(1)/2-1 S(1) line ratio ranges from $\sim$7 to $\sim$10 suggesting the presence of shock interactions. Our findings can be explained by the interaction of a jet/bullet ejected from the central star with the surrounding asymptotic giant branch material. The strongest H$_2$ line, v=1-0 S(1), is also detected in several low-ionization knots located at the periphery of the elliptical planetary nebula NGC~7662, but only four of these knots are detected in the H$_2$ v=2-1 S(1) line. These four knots exhibit an H$_2$ line ratio between 2 and 3.5, which suggests that the emission is caused by the UV	ionizing flux of the central star. Our data confirms the presence of H$_2$ gas in both fast- and slow-moving low-ionization knots, which has only been confirmed before in the nearby Helix nebula and Hu 1-2. Overall, the low-ionization structures of planetary nebulae are found to share similar traits to photodissociation regions.

\noindent
\end{abstract}


\section{Introduction}

Optical imaging surveys of planetary nebulae (PNe) have revealed that a significant fraction of PNe possess,
in addition to large-scale structures such as rims, shells and haloes, various small-scale structures (e.g. Balick 1987; Balick et al. 1998; 
Corradi et al. 1996; Gon\c{c}alves et al. 2001). In these small-scale structures low-ionization emission lines such as \oxygeni, \nitrogen, \oxygenii\ and \sulfurt\ (low-ionization emission line structures, hereafter LISs) are more prominent than the large scale structures. Moreover, they have a variety of shapes and forms such as knots, jets and filaments (e.g. Gon\c{c}alves et al. 2001, and references therein) and cover a wide range of expansion velocities from a few ten of km s$^{-1}$ up to a few hundreds of km s$^{-1}$, indicating a variety of formation and excitation mechanisms.

Shock interactions play a crucial role in the excitation of LISs and provide a plausible explanation for the enhancement of the low-ionization emission lines like \nitrogen, given that there are no chemical abundances variations (e.g. N and O) between the LISs and the surrounding medium (e.g. Gon\c{c}alves et al. 2006; Akras \& Gon\c{c}alves 2016). By studying a sample of five Galactic PNe with LISs and comparing them with theoretical predictions, Akras \& Gon\c{c}alves (2016) demonstrated that the excitation mechanism of LISs is via a combination shocks and UV-photons emitted from the central star. The contribution of each mechanism depends on a number of physical parameters such as the 
distance of LISs to the central star, the stellar parameters (T$_{\rm{eff}}$, L$_{\odot}$), the expansion velocity and the density. 
The same  concept of a mixture of UV-photons produced by shocks and UV-photons emitted from the central star, is also used to explain
the bright low-ionization lines in highly evolved PNe (Akras et al. 2016).

The physical properties of LISs such as electron temperature (T$_{\rm{e}}$), electron density (n$_{\rm{e}}$) and chemical abundances 
are also of great importance for understanding these structures. They were first studied in detail by Balick and collaborators 
(Balick et al. 1993, 1994, 1998). The interpretation given by these authros of the \nitrogen\ emission line enhancement was the likely 
overabundance of nitrogen compared with the surrounding medium. Later, Gon\c{c}alves and collaborators (Gon\c{c}alves et al. 2003, 2009) proved 
that this is not a necessary condition for explaining the enhancement of the LISs. 
Akras \& Gon\c{c}alves (2016) recently shown that the enhancement of low-ionization lines in LISs is the result of their different degree of ionization compared with the rest of the nebula.

Also of importance is the difference in n$_{\rm{e}}$ between the LISs and the other nebular components, as the former are usually 
found to be less dense than the surrounding medium (Gon\c{c}alves et al. 2009). This observational result is inconsistent with the formation models of LISs (e.g. shock models, Dopita 1997; stagnation model, Steffen et al. 2001; magneto--hydrodynamic interacting winds model, 
Garc\'{i}a--Segura et al. 2005; binary systems, Soker \& Livio 1994), which predict that LISs will be denser than the surrounding medium. 
An explanation for this discrepancy between the theoretical models and observations could be that the models 
refer to the total density (dust, atomic and molecular) and not only to the n$_{\rm{e}}$, which corresponds to the 
ionized fraction of the gas. Therefore, a possible solution may be that LISs are also made of non--ionized gas (molecular, atomic), as 
proposed by Gon\c{c}alves et al. (2009). This hypothesis is confirmed for a cometary knot in the well studied PN Helix, 
for which the molecular mass is found to be substantially higher than the ionized mass  (Huggins et al 2002; Matsuura et al. 2009). 
More recently, Fang et al. (2015) detected H$_2$ emission from the knots of Hu 1-2. Moreover, high-angular resolution images of NGC~2346 in 
the H$_2$ v=1-0 line revealed that the equatorial H$_2$ region of the nebula is fragmented into knots and filaments (Manchado et al. 2015).

H$_2$ emission is commonly detected in PNe (e.g. Latter et al. 1995; Kastner et al. 1996; Guerrero et al. 2000; 
Arias et al. 2001; Ramos--Larios et al. 2012; Marquez-Lugo et al. 2015) and proto--PNe (Sahai et al. 1998; Hrivnak et al. 2008) mainly 
via the bright ro--vibrational v=1--0 transition line of H$_2$ at 2.122 $\mu$m. The detection rate of H$_2$ emission has been found 
to be higher in bipolar PNe than in other morphological types (e.g. round and/or elliptical). This is known as Gatley's rule (Kastner et al. 1994). However, there are some round and elliptical PNe in which H$_2$ emission has been detected (e.g. the round NGC 6781, Phillips et al. 2011; the elliptical NGC 6720, Latter et al. 1995; NGC 7048, Ramos--Larios et al. 2013; see also Marquez-Lugo et al. 2013). These round and elliptical PNe have central stars with effective temperatures (T$_{\rm{eff}}$) $>$100~kK. 
Therefore, the high detection rate of H$_2$ in bipolar PNe may be associated with the high T$_{\rm{eff}}$ of their central stars (Phillips 2006; Aleman \& Gruenwald 2004, 2011).

A comparison of H$_2$ with optical emission line images (e.g. \nitrogen\ and \oxygeni) has revealed very similar morphologies (Guerrero et al. 2000; Bohigas 2001), which suggests that the two emissions originate from the same regions. 
Theoretical works by Aleman and collaborators (Aleman \& Gruenwald 2011; Aleman et al. 2011) have shown that the rovibrational emission of H$_2$ may be important, not only in the photo-dissociation regions (PDRs) but also in the transition zone between the ionized and the neutral  gas in PNe, a zone containing gas of intermediate ionization degree (see Fig. 4 of Aleman \& Gruenwald 2011). Moreover, these authors have shown that the distance from the central star, or equivalently the intensity of the UV radiation field, is a crucial parameter regarding the ionization structure of the knots. Inside the ionization front, knots with high optical depths will exhibit strong H$_2$, \nitrogen\ and \oxygeni\ emission lines, whereas those beyond the ionization front will emit only in the H$_2$ and \oxygeni\ lines (Aleman et al. 2011). Overall, the \oxygeni\ emission seems to be a better indicator 
for the presence of H$_2$ than the \nitrogen\ emission. However, in PNe low-ionization knots, which are dense structures, 
either \nitrogen\ or \oxygeni\ emission line would imply the presence of H$_2$ gas. All cometary knots with \nitrogen\ emission in the Helix nebula 
also exhibit H$_2$ emission (Matsuura et al. 2009). Moreover, it is worth mentioning that an empirical linear relation between the 
fluxes of the \oxygeni\ 6300~\AA\ emission line and the H$_2$ v=1-0 for Galactic PNe was proposed many years ago by Reay et al. (1988).

In this paper, we present the detection of H$_2$ emission from fast- and slow-moving low-ionization knots in two Galactic PNe, namely K~4-47 
and NGC~7662. The observations and data analysis are described in Section 2. The results of this work are discussed in Section 3, 
and the conclusion is presented in Section 4.   

\section{Observations}

\begin{table}
\caption{Observations log} 
\scriptsize{
\begin{tabular}{lcccccc}
\hline
\hline
Object & \multicolumn{1}{c}{Filter} & \multicolumn{1}{c}{$\rm{\lambda_c}$} &
\multicolumn{1}{c}{$\rm{\Delta\lambda}$} & \multicolumn{1}{c}{Time} & \multicolumn{1}{c}{No. of frames} \\
        &           &  ($\micron$) & ($\micron$) & (s) & \\
\hline
K~4-47   &  K-cont-1    & 2.0975 & 0.0275 &  90  & 9 \\
NGC~7662 &  K-cont-1    & 2.0975 & 0.0275 &  115 & 9 \\
K~4-47   &  H$_2$ v=1-0 S(1) & 2.1239 & 0.0261 &  90  & 9 \\
NGC~7662 &  H$_2$ v=1-0 S(1) & 2.1239 & 0.0261 &  115 & 9 \\
K~4-47   &  Brackett$\rm{\gamma}$  & 2.1686 & 0.0295 &  60  & 3 \\
NGC~7662 &  Brackett$\rm{\gamma}$  & 2.1686 & 0.0295 &  75  & 3 \\ 
K~4-47   &  H$_2$ v=2-1 S(1) & 2.2465 & 0.0301 &  155 & 21 \\
NGC~7662 &  H$_2$ v=2-1 S(1) & 2.2465 & 0.0301 &  190 & 21$^{\dag}$ \\
K~4-47   &  K-cont-2    & 2.2718 & 0.0352 &  155 & 21 \\
NGC~7662 &  K-cont-2    & 2.2718 & 0.0352 &  190 & 21$^{\dag}$ \\              
\hline
\end{tabular}
}
\begin{flushleft}
\scriptsize{
$^{\dag}$ For these observations only 14 out of 21 frames could be used for analysis. \\
}
\end{flushleft}
\end{table}

\begin{figure*}
\centering
\includegraphics[scale=0.349]{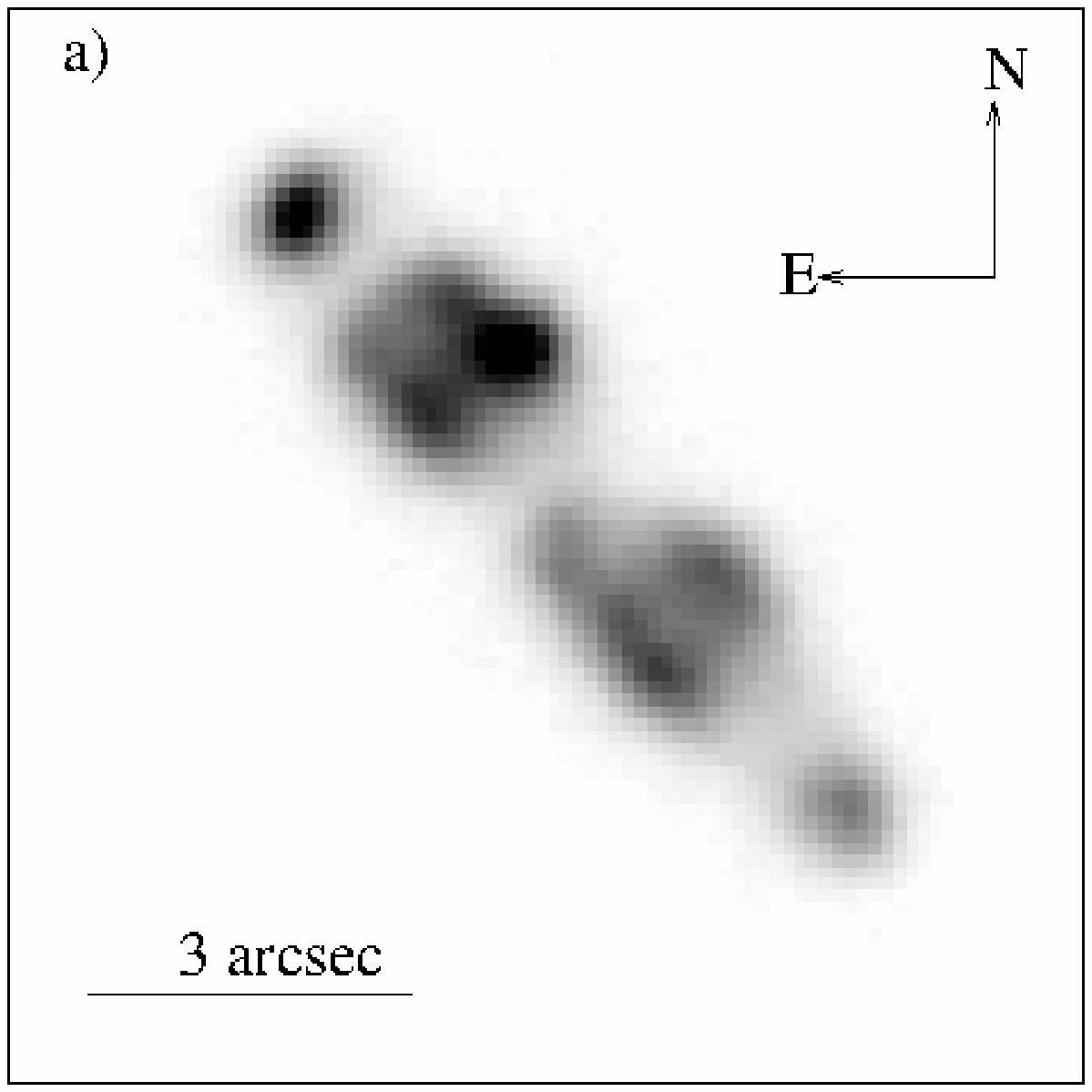}
\includegraphics[scale=0.349]{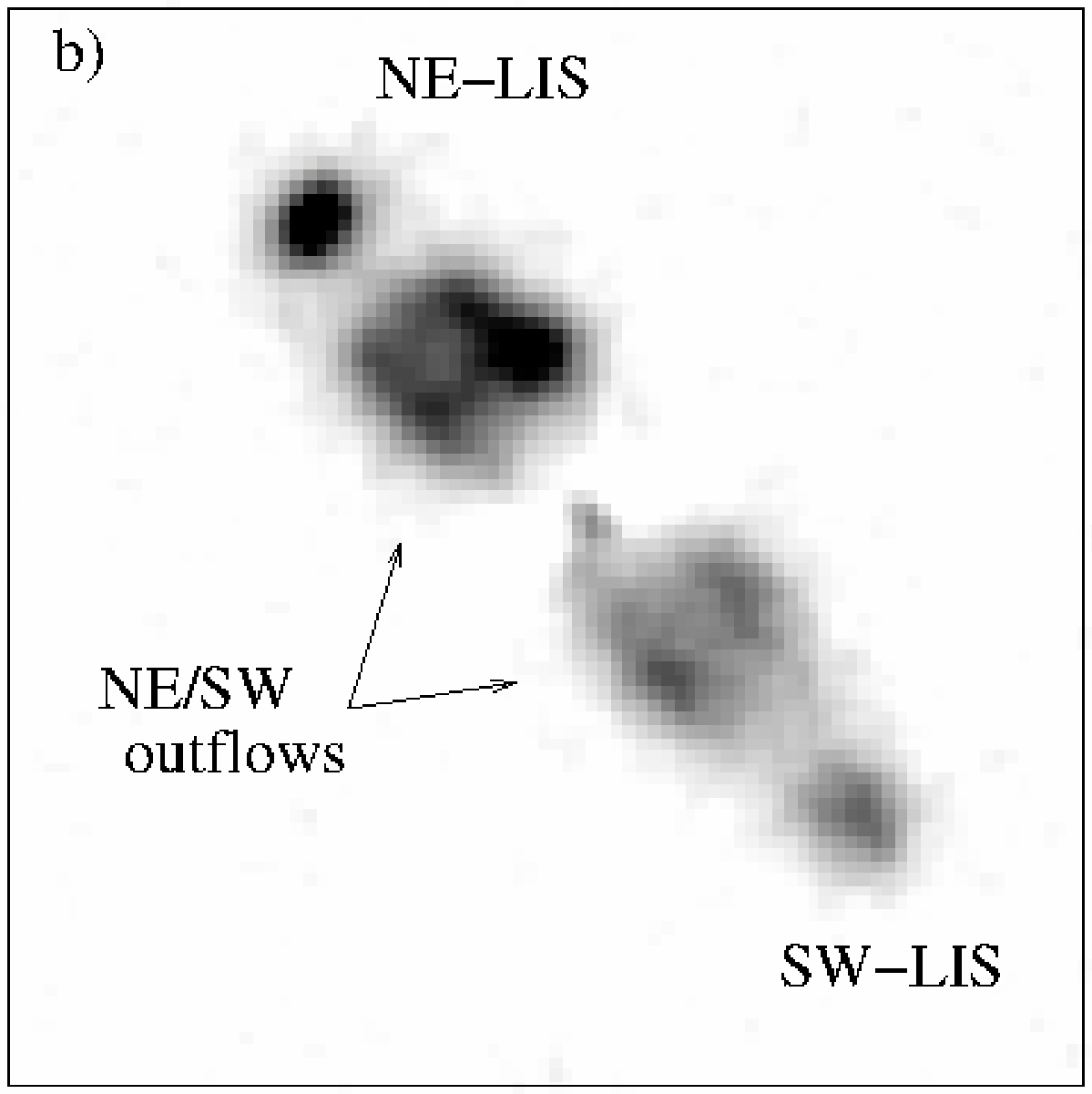}
\includegraphics[scale=0.349]{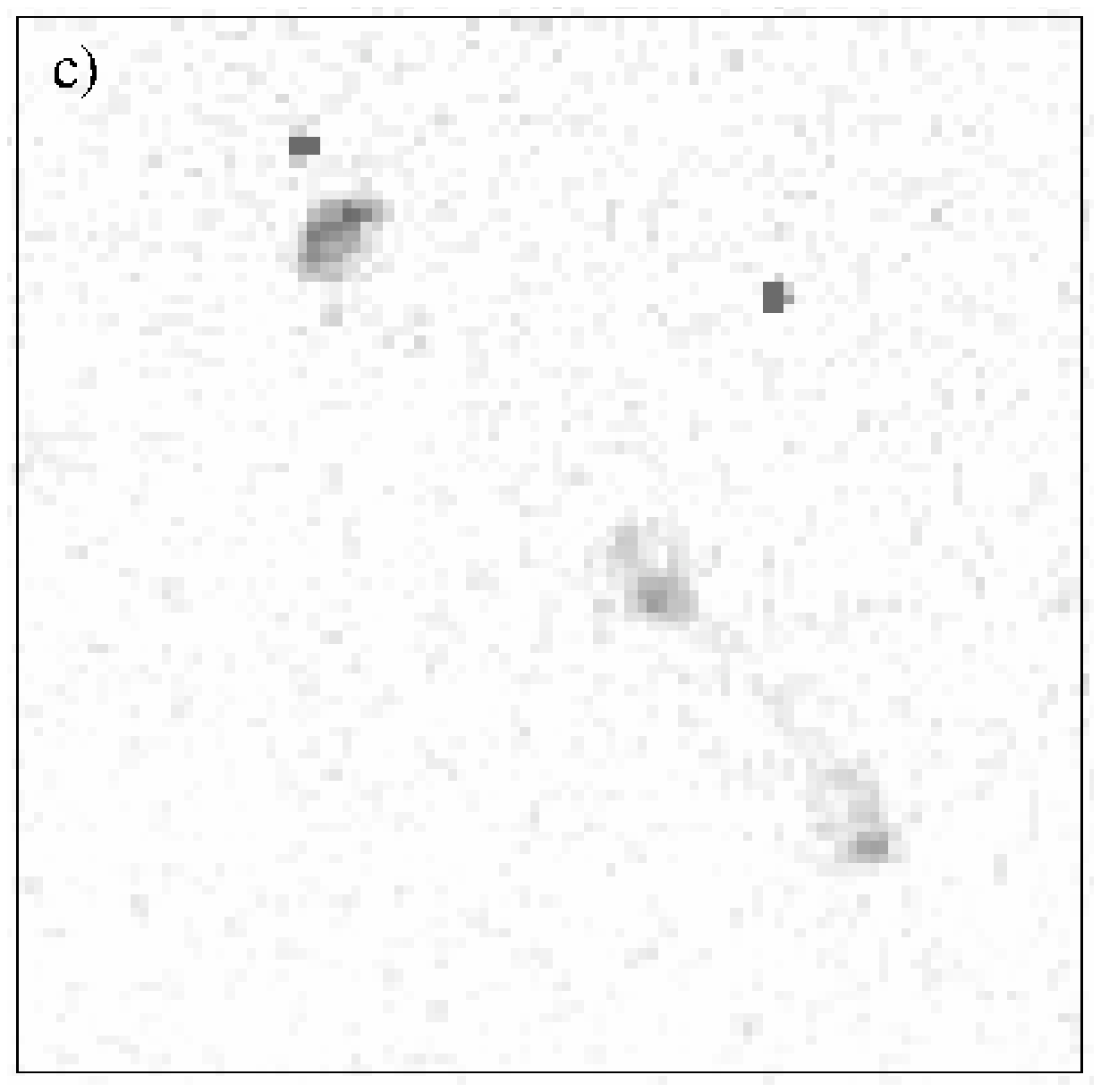}
\includegraphics[scale=0.350]{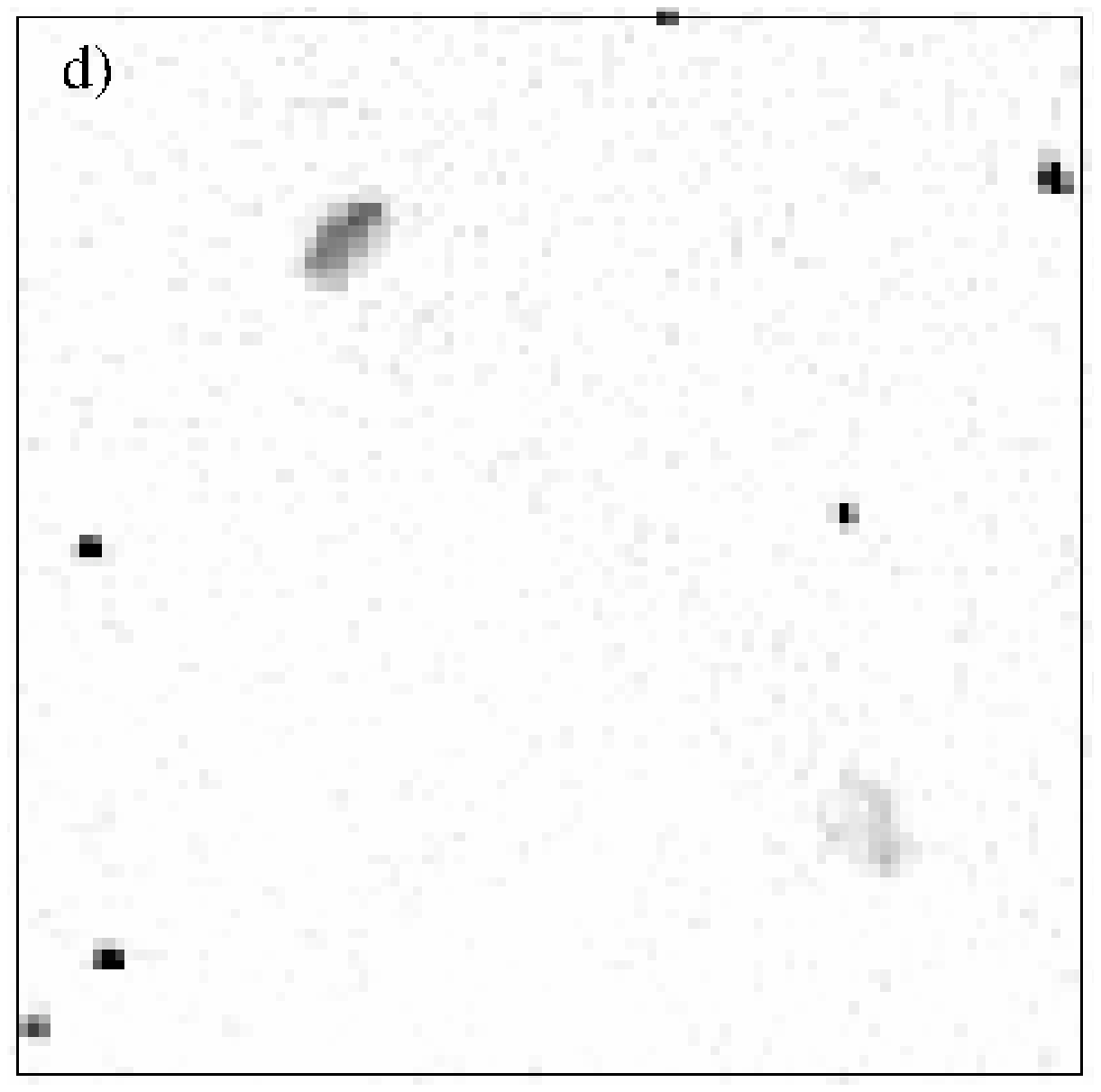}
\caption[]{Emission line images of K 4-47: (a) NIRI $\rm{H_2}$ v=1-0 S(1); (b)  NIRI $\rm{H_2}$ v=2-1 S(1); (c)  {\it HST} \nitrogen\ $\lambda$6584; and (d)  {\it HST} \oxygeni\ $\lambda$6300. The size of the boxes is 20$\times$20~arcsec$^2$. NE-LIS and SW-LIS correspond to Knot1 and Knot2, respectively, in Corradi et al. (2000) and Gon\c calves et al. (2004).} 
\label{fig1}
\end{figure*}

High-angular-resolution, near-IR narrow-band H$_2$ (namely v=1-0 S(1) at 2.122~$\mu$m, v=2-1 S(1) at 2.248~$\mu$m and Br$\gamma$ at 
2.168$\mu$m) images of two Galactic PNe with LISs, namely K~4-47 and NGC~7662, were obtained using the Near InfraRed Imager and Spectrometer (NIRI) on the 8~m~Gemini-North  telescope, at Mauna Kea in Hawaii. The observations were carried out 
in the service observing mode on 2014 August 6$^{\rm{th}}$ and 2014 October 13$^{\rm{th}}$ (Program IDs: GN2014B-Q43 and GN2014B-Q1). 

The observations were carried out in the f/6 configuration (pixel scale = 0.117~arcsec and fields of view of 120~arcsec). 
The narrow-band G0216, G0218 and G0220 filters, centered at 2.1239, 2.1686 and 2.2465$\mu$m, respectively, 
were applied to isolate the H$_2$ v=1-0, Br$\gamma$ and H$_2$ v=2-1 lines, respectively. Continuum images adjacent to the 
above lines were also obtained, using the corresponding filters centered at 2.0975~$\mu$m (G0217) and 2.2718~$\mu$m (G0219), which allowed us 
to subtract the continuum contribution from the line emission. Details about the  observations are given in Table~1.

The exposure times were estimated using the empirical relation between the fluxes of \oxygeni\ 6300~\AA\ and H$_2$ v=1-0 lines, as
reported by Reay et al. (1988). These times (total integration times) and the number of individual frames are summarized in Table~1. GCAL flats frames were also obtained for the correction of thermal emission, 
dark current and hot pixels, in accordance with the Gemini baseline calibrations. For the correction of sky background as well as bad pixels, the target was dithered across the CCD detector. In order to flux-calibrate the images, we also observed 
the infrared standard stars FS108 (RA: 03 01 09.8, Dec: +46 58 47.8) and FS 104 (RA: 01 04 59.6, Dec: +41 06 31.3) (Hawarden et al. 2001).
 
The images were reduced using the tasks within the Gemini IRAF package for NIRI ({\it Nprepapre, Nisky, Niflat, Nireduce and Imcoadd}). 
Before starting the data reduction, the python routines CLEARIR.py 
and NIRLIN.py were used in order to correct the vertical stripping and the non-linearity of the detector, which affected the raw data. 
The reduced H$_2$ images are presented in Figure 1.  The continuum subtracted Br$\gamma$ images of K~4-47 and NGC~7662 are not shown here because either they are too noisy (K~4-47) or the morphology is very similar to the optical {\it HST} image (NGC~7662). Br$\gamma$ emission in K~4-47 
probably originates from the inner core of the nebula, and is very weak in the bipolar outflows and knots. Fig. 2 shows the Br$\gamma$ emission image of K~4-47 (not continuum subtracted), overlaid with the contours of the H$_2$ v=1-0 emission line. 

The 3$\rm{\sigma}$ detection limits of the H$_2$ v=1-0 S(1) and Br$\gamma$ line images are 2.7$\times$10$^{-16}$~erg cm$^{-2}$ s$^{-1}$ arcsec$^{-2}$ and 9$\times$10$^{-16}$~erg cm$^{-2}$ s$^{-1}$ arcsec$^{-2}$, respectively. For the H$_2$ v=2-1 S(1) line images the 3$\rm{\sigma}$ are 1.1$\times$10$^{-16}$~erg cm$^{-2}$ s$^{-1}$ arcsec$^{-2}$, for the K~4-47, and 2.5$\times$10$^{-16}$~erg cm$^{-2}$ s$^{-1}$ arcsec$^{-2}$, for the NGC~7662, due to the use of different number of frames (see Table~1).

\begin{table*}
\begin{center}
\caption{Emission line fluxes for K~4-47, in units of 10$^{-15}$ erg~s$^{-1}$~cm$^{-2}$. R(H$_2$) = H$_2$ v=1-0/v=1-2 and R(Br$\gamma$) = H$_2$ v=1-0/Br$\gamma$. Numbers in parenthesis correspond to the errors. Fluxes without errors correspond to the 3$\rm{\sigma}$ upper limits, whereas ratios without errors correspond to lower limits. Radii are in units of arcseconds.} 
\begin{tabular}{lcccccccc}
\hline
\hline
\noalign{\smallskip}
Name & \multicolumn{1}{c}{RA} & \multicolumn{1}{c}{Dec} &
\multicolumn{1}{c}{H$_2$ v=1-0} & \multicolumn{1}{c}{H$_2$ v=2-1} & \multicolumn{1}{c}{Br$\rm{\gamma}$}
& R(H$_2$) & R(Br$\rm{\gamma)}$ & \multicolumn{1}{c}{Radius} \\
\hline
NE-LIS      & 04:20:45.5 & 56:18:16.2 &  6.82 (0.31) & 0.96 (0.05) & 0.96 &  7.12 (0.43) & 7.05 & 0.585\\
SW-LIS      & 04:20:44.9 & 56:18:10.6 &  4.29 (0.19) & 0.64 (0.04) & 0.96 &  6.71 (0.39) & 4.47 & 0.585\\
NE-outflow  & 04:20:45.4 & 56:18:14.8 & 21.89 (0.45) & 2.57 (0.06) & 2.47 &  8.51 (0.23) & 8.86 & 0.936\\
SW-outflow  & 04:20:45.1 & 56:18:12.4 & 18.38 (0.31) & 1.83 (0.05) & 2.47 & 10.05 (0.24) & 7.45 & 0.936\\
\hline
\end{tabular}
\end{center}
\end{table*}

\begin{figure}
\centering
\includegraphics[scale=0.50]{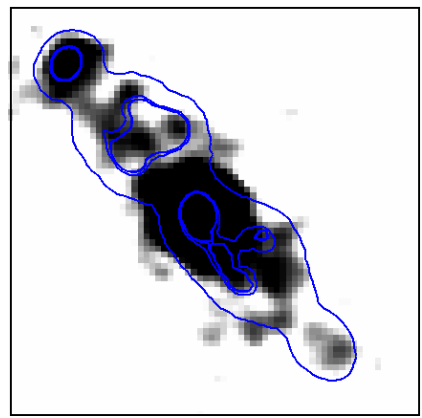}
\caption[]{Contours of the $\rm{H_2}$ v=1-0 S(1) line, overlaid on the Br-$\rm{\gamma}$ image of K~4-47 (not continuum subtracted). The box size is 9$\times$9~arcsec$^2$. North is up and east is to the left.} 
\label{fig2}
\end{figure}

\section{Results}
\subsection{The highly collimated PN K~4-47}

Optical images of K-4-47 have revealed a jet-like, highly collimated structure with a pair of fast-moving knots (Corradi et al. 2000). 
These knots (where NE-LIS and SW-LIS correspond to Knot1 and Knot2, respectively, in Corradi et al. (2000) and Gon\c calves et al. (2004)) 
exhibit very strong \nitrogen\ and \oxygeni\ emission lines (\nitrogen\ 5200/H$\alpha$=367 (NE-LIS) and 410 (SW-LIS) 
and \oxygeniii\ 6300/H$\alpha$=427 (NE-LIS) and 305 (SW-LIS), probably implying a strong shock interaction, 
which is consistent with their high projected expansion velocities of $\sim$100~\kms\ (Corradi et al. 2000). 
The {\it HST} \nitrogen\ $\lambda$6584 and \oxygeni\ $\lambda$6300 images display strong emission in these knots (see Fig.~1, 
panels c and d, respectively). The distance of the knots from the central star -- 3.9~arcsec (NE-LIS) and 3.45~arcsec (SW-LIS) -- are slightly different from the values measured by Corradi et al. (2000). However, the distances measured in the optical {\it HST} images agree better 
with Corradi et al.'s values.

After analyzing a number of shock models, Gon\c{c}alves et al. (2004) came to the conclusion that in order to reproduce the line intensities (or the excitation of the LISs) of this nebula, expansion velocities up to 250-300~$\kms$, in conjunction with the UV-photons from the central star, are required. Although, the true nature of K~4-47 still remains unknown (is it really a PN, or it could be a symbiotic?), the mechanism 
responsible for the prominent low-ionization lines is at least partially a result of shock interactions.   

From the point of view of molecular chemistry, K~4-47 is a very interesting object, in which many molecules species have been found. Huggins et al. (2005) first detected strong CO emission, and few years later Edwards et al. (2014) detected emission in HCO$^{+}$ and CS. Very recently, HCN emission was also detected in this object (Schmidt and Ziurys 2016). It should be noted, however, that the H$_2$ densities of K~4-47 calculated by Edwards et al. (2014) and Schmidt and Ziurys (2016) differ by a factor of 10. This discrepancy may be a result of the assumption of Schmidt and Ziurys (2016) 
that the molecules exist in dense clumps. 

We present here the first narrow-band near-IR H$_2$ images of K 4-47. These images display a remarkable highly collimated bipolar structure, 
with the emission arising from the walls of the bipolar outflows, resembling a hollow conical structure, and a pair of low-ionization 
knots apparent at the tips of these outflows (Fig.~1). The detection of H$_2$ emission in the outflows of K~4-47 proves that the 
assumption of Schmidt and Ziurys (2016) is not suitable for this object. Furthermore, none of these structures is detected in our Br$\gamma$ 
images. We thus obtain an upper limits of the Br$\gamma$ line, in order perform further relevant analysis in Section 4. It should be 
noted that Lumsden et al. (2001) reported a Br$\gamma$ flux of 4.29$\times$10$^{-15}$ erg~s$^{-1}$~cm$^{-2}$ integrated 
along the slit (2.4 arcsec$\times$9.6 arcsec). This emission likely originates mainly from the core of K~4-47. From an inner region 
of 1.35~arcsec$^2$, we calculate F(Br$\gamma$)=(3.3$\pm$0.29)$\times$10$^{-15}$ erg~s$^{-1}$~cm$^{-2}$.

\begin{table*}
\begin{center}
\caption{Emission line fluxes for the low-ionization structures of NGC~7662, in units of 10$^{-16}$ erg~s$^{-1}$~cm$^{-2}$. 
R(H$_2$) = H$_2$ v=1-0/v=1-2 and R(Br$\gamma$) = H$_2$ v=1-0/Br$\gamma$. 
Numbers in parenthesis are the errors of the fluxes. Flux without errors correspond to the 3$\rm{\sigma}$ upper limits, whereas the ratios without errors correspond to lower limits. Radii are in units of arcseconds. }
\begin{tabular}{lcccccccc}
\hline
\hline
Name & \multicolumn{1}{c}{RA} & \multicolumn{1}{c}{Dec} &
\multicolumn{1}{c}{H$_2$ v=1-0} & \multicolumn{1}{c}{H$_2$ v=2-1} &\multicolumn{1}{c}{Br$\gamma$} 
& R(H$_2$) & R(Br$\gamma$)  & \multicolumn{1}{c}{radius} \\
\hline
LIS1  & 23:25:54.7 & 42:32:21.5 & 1.88(0.17) &  1.72& 6.18        & 1.09 & 0.30 & 0.468\\
LIS2  & 23:25:54.5 & 42:32:21.6 & 1.96(0.18) &  1.72             & 6.18        & 1.14       & 0.31 & 0.468\\
LIS3  & 23:25:54.6 & 42:32:17.5 & 1.05(0.11) &  0.96             & 6.62 (0.15) & 1.10       & 0.16 (0.02) & 0.351\\
LIS4  & 23:25:54.6 & 42:32:16.2 & 4.23(0.35) &  1.72             & 13.8 (0.41) & 2.46       & 0.31 (0.03) & 0.468\\
LIS5  & 23:25:54.4 & 42:32:14.8 & 5.14(0.39) &  1.72             & 21.9 (0.33) & 2.99       & 0.23 (0.02) & 0.468\\
LIS6  & 23:25:54.7 & 42:32:15.4 & 1.38(0.13) &  0.96             & 7.68 (0.15) & 1.44       & 0.18 (0.02) & 0.351\\
LIS7  & 23:25:54.8 & 42:32:13.8 & 8.63(0.31) &  2.89(0.15)       & 37.1 (0.69) & 2.99(0.19) & 0.23 (0.02) & 0.702\\
LIS8  & 23:25:54.9 & 42:32:09.7 & 12.6(0.51) &  4.14(0.21)       & 62.8 (0.79) & 3.04(0.20) & 0.20 (0.02) & 0.702\\
LIS9  & 23:25:54.9 & 42:32:09.1 & 3.95(0.24) &  1.72             & 24.5 (0.22) & 2.30       & 0.16 (0.02) & 0.468\\
LIS10$^a$ & 23:25:54.4 & 42:32:10.9 & 1.61(0.26) &  1.72         & 4.94 (0.37) & 0.94       & 0.33 (0.05) & 0.468\\
LIS11$^a$ & 23:25:54.6 & 42:32:09.6 & 1.44(0.29) &  0.96         & 3.94 (0.34) & 1.51       & 0.37 (0.06) & 0.351\\
LIS12$^a$ & 23:25:54.5 & 42:32:07.6 & 1.73(0.27) &  0.96         & 3.85 (0.41) & 1.80       & 0.44 (0.05) & 0.351\\
LIS13 & 23:25:54.7 & 42:32:06.7 & 1.81(0.29) &  0.96             & 21.2 (0.31) & 1.89       & 0.09 (0.02) & 0.351\\
LIS14 & 23:25:54.8 & 42:32:05.4 & 1.59(0.13) &  0.96             & 16.3 (0.19) & 1.66       & 0.10 (0.02) & 0.351\\
LIS15 & 23:25:54.9 & 42:32:05.5 & 1.39(0.11) &  0.96             & 13.8 (0.21) & 1.45       & 0.10 (0.02) & 0.351\\
LIS16 & 23:25:54.9 & 42:32:01.2 & 1.69(0.12) &  0.96             & 28.4 (0.39) & 1.76       & 0.06 (0.02) & 0.351\\
LIS17 & 23:25:54.7 & 42:32:00.3 & 1.59(0.12) &  0.96             & 22.9 (0.32) & 1.66       & 0.07 (0.02) & 0.351\\
LIS18 & 23:25:54.6 & 42:31:58.7 & 1.46(0.13) &  0.96             & 16.8 (0.32) & 1.52       & 0.09 (0.02) & 0.351\\
LIS19a& 23:25:53.7 & 42:31:58.2 & 5.62(0.35) &  1.72             & 94.7 (1.21) & 3.27       & 0.06 (0.02) & 0.468\\
LIS19b& 23:25:53.7 & 42:31:57.3 & 4.95(0.37) &  1.72             & 80.9 (0.48) & 2.88       & 0.06 (0.03) & 0.468\\
LIS20 & 23:25:53.7 & 42:31:52.0 & 14.3(0.84) &  4.41(0.16)       & 28.4 (0.79) & 3.24(0.22) & 0.50 (0.04) & 0.702\\
LIS21 & 23:25:53.6 & 42:31:50.4 & 5.38(0.46) &  2.68             & 9.67        & 2.01       & 0.55        & 0.585\\
LIS22 & 23:25:53.5 & 42:31:48.2 & 0.96(0.09) &  0.96             & 3.47        & 1.00       & 0.28        & 0.351\\
LIS23 & 23:25:53.5 & 42:31:47.3 & 1.07(0.11) &  0.96             & 3.47        & 1.12       & 0.31        & 0.351\\
LIS24 & 23:25:53.2 & 42:32:02.9 & 5.72(0.52) &  1.72             & 37.2 (1.24) & 3.33       & 0.16 (0.02) & 0.468\\
LIS25 & 23:25:52.9 & 42:32:01.5 & 2.26(0.21) &  0.96             & 6.51 (0.19) & 2.36       & 0.35 (0.03) & 0.351\\
LIS26 & 23:25:52.9 & 42:31:57.6 & 11.2(0.54) &  5.14(0.22)       & 45.7 (0.98) & 2.18(0.15) & 0.25 (0.02) & 0.702\\
LIS27 & 23:25:53.2 & 42:31:54.8 & 6.09(0.33) &  3.87             & 25.9 (0.68) & 1.58       & 0.24 (0.02) & 0.702\\
LIS28 & 23:25:54.2 & 42:32:18.8 & 0.72(0.14) &  0.96             & 10.3 (0.31) & 0.75       & 0.07 (0.02) & 0.351\\
\hline
\end{tabular}
\end{center}
\begin{flushleft}
$^a$ For these LISs the residual emission of the rim has been estimated and subtracted.\\
\end{flushleft}
\end{table*}

\subsection{The elliptical planetary nebula NGC 7662}

NGC~7662 is an elliptical PN that possesses more than two dozen of LISs (knots and one jet-like feature at the southern part), mainly embedded  in the outer shell. The expansion velocities of these LISs are of the order of 30~$\kms$, except for the southern jet-like feature that has V$_{\rm{exp}}$=70~$\kms$ (Perinotto et al. 2004). All these structures are easily discerned in the \nitrogen\ image (see Guerrero et al. 2004; Gon\c calves et al. 2009) or the {\it HST} \nitrogen / \oxygeniii\ line ratio image, as shown in Fig.~3. Spectroscopic data from these structures have also revealed strong \oxygeni\ $\lambda$6300 and \nitrogen\ $\lambda$6584 emission lines (Perinotto et al. 2004; Gon\c{c}alves et al. 2009). Despite the fact that the former lines are important indicators of shock interaction, most of the LISs in NGC~7662 have low expansion velocities, similar to the overall nebular expansion velocity (Perinotto et al. 2004). Thus, we can ask which mechanism, if not shocks, is responsible for the enhancement of the low-ionization emission lines?

From the point of view of the optical emission line ratios, it cannot be deduced whether the LISs of NGC~7662 are shock-excited or 
photoionized regions. Although the LISs of NGC 7662 are found to be closer to the regime occupied by fast low-ionization structure (FLIERs) than the regime of rims and shells in the diagnostic diagrams in Raga et al. (2009), they do lie in a very narrow transition zone between the two regimes (see Fig.~6 in Gon\c{c}alves et al. 2009). Therefore, both mechanisms (shocks and UV-radiation) should contribute to the overall excitation of these structures.

\begin{figure*}
\centering
\includegraphics[scale=0.325]{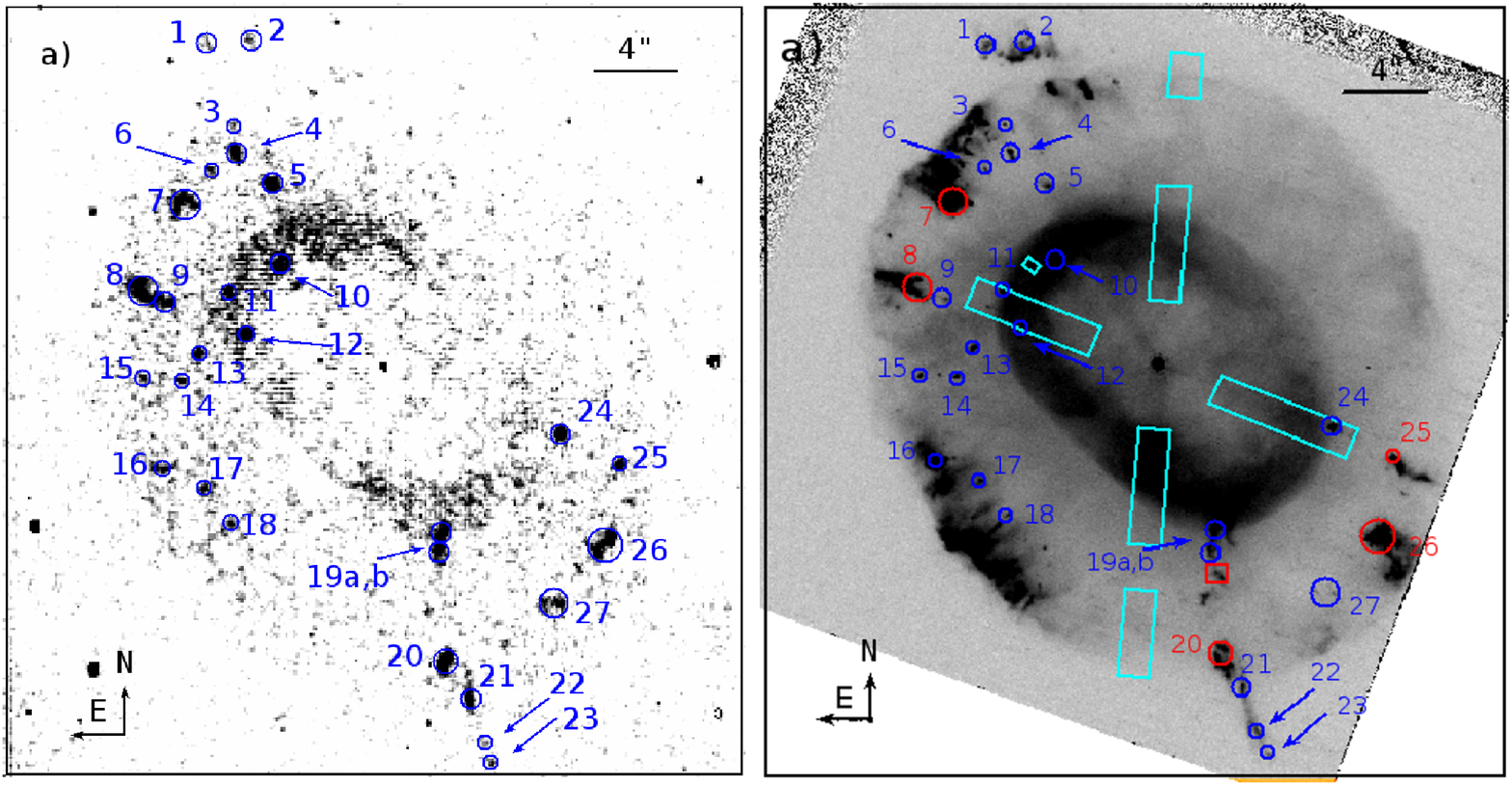}
\caption[]{Emission line images of NGC~7662: a)  NIRI $\rm{H_2}$ v=1-0 S(1) continuum-subtracted; and 
b) {\it HST} \nitrogen / \oxygeniii\ line ratio image. The circles (in both panels) correspond to the LISs 
detected in H$_2$. The red circles and cyan boxes in panel (b) indicate LISs and nebular regions with available 
optical spectra (Perinotto et al. 2004, Gon\c{c}alves et al. 2009). Both structures have been used in the optical diagnostic diagram (Fig.~5). The red box indicates a LIS with available optical spectrum but not detected in 
H$_2$ emission. The box size is 35$\times$37~arcsec$^2$. 
} 
\label{fig3}
\end{figure*} 

We present here deep near-IR images of this nebula. The H$_2$ v=1-0 S(1) line is detected in several optically 
identified LISs (Fig.~3, right panel), but only four of them (LIS7, LIS8, LIS20 and LIS26) have H$_2$ v=2-1 S(1) emission, 
higher than the 2$\rm{\sigma}$ level. In Table~3 we list the details for each individual LIS in NGC~7662. From the four LISs with both H$_2$ lines detected, we estimated R(H$_2$) -- the H$_2$ v=1-0/v=1-2 line ratio -- to be between 2 and 3.5, whereas for the others we provide lower limits (Table~3).

The very hot and luminous central star of NGC 7662 (T$_{\rm{eff}}$=95kK, log(L/L$_{\odot}$)=3.42; Henry et al. 2015) emits a large number 
of high-energy UV-photons (high photo-ionization rate), which, in conjunction with the low expansion velocities of the LISs, suggests 
that the LISs in this nebula are photoionized predominantly by the radiation field of the central star (Raga et al. 2008; Gon\c{c}alves et al. 2009; Aleman et al. 2011; Akras \& Gon\c{c}alves 2016). Furthermore, all the LISs in this PN are found to be Br$\gamma$-bright sources (R(Br$\gamma$)$<$1) with a low R(H$_2$) ratio, which seems to be a common characteristic of photoionized regions (Marquez-Lugo et al. 2015) 
{\it (See the Sect.~4 for a more complete discussion)}.

\section{Discussion} 
Motivated by the paradoxical electron density contrast of LISs and their surrounding medium versus that of the formation models, the H$_2$ and Br$\gamma$ line images of two PNe have been used here for a more detailed study of LISs.

K~4-47 shows a pair of fast-moving knots at the tips of the collimated bipolar outflows. Strong H$_2$ lines were detected from the knots and the walls of the bipolar outflows. The R(H$_2$) ratio is found to vary from $\sim$7 to 10.
These values indicate that the excitation of H$_2$ gas is powered by shocks, which is consistent with the high expansion velocities of the knots (100~\kms; Corradi et al. 2000). The strong neutral emission lines found in the knots (\nitrogena\ and \oxygeni) have also been attributed to shock interactions (Gon\c{c}alves et al. 2004).  

One way to investigate the excitation mechanism of the H$_2$ gas is by calculating the R(H$_2$) line ratio. 
In principle, shock-excited regions exhibit R(H$_2$)$>$10, whereas photoionized regions can cover a wider range of R(H$_2$) 
values, from 2 up to 10, depending on the hydrogen gas density and UV radiation intensity. In particular, low density structures exhibit R(H$_2$) values close to 3, in contrast to high density structures ($>$10$^5$~cm$^{-3}$) that 
can have R(H$_2$) values up to 10. This make it hard to deduce whether the emission lines are produced in photoionized or shock-excited regions 
(see Black \& van Dishoeck 1987; Sternberg \& Dalgarno 1989; Burton et al. 1990).
K~4-47 has R(H$_2$) values between {\bf $\sim$7}(LISs) and 9-10 (bipolar outflows). 
These R(H$_2$) values imply that the excitation mechanism may not be the same, for the two kinds of structures, owing to the substantial 
difference in their expansion velocities. 
Our results suggest that K~4-47 is predominantly excited by shocks in agreement with the previous studies (Corradi et al. 2000; Lumsden et al. 2001). However, a contribution of UV-photons from the central star cannot be ruled out (Gon\c{c}alves et al. 2004; Akras \& Gon\c{c}alves 2016). The non-detection of Br$\gamma$ emission from the bipolar outflows and knots as well as the faint optical emission (Corradi et al. 2000, Gon\c{c}alves et al. 2004) imply a very low contribution from the UV-photons to the excitation of K~4-47. 

However, the R(H$_2$) ratio alone is not enough to discrimine the shock-excited from photoionized regions, as shown by, for example, Marquez-Lugo et al. 2015. These authors demonstrated that shock-excited regions are also brighter in H$_2$ than in Br$\gamma$. In Table~2, we list the 
3$\rm{\sigma}$ Br$\gamma$ detection as well as the lower limit of the R(Br$\gamma$)= H$_2$ v=1-0/Br$\gamma$ line ratio. The knots and bipolar outflows of K~4-47 are found to exhibit R(Br$\gamma$)$>$5, thus providing further evidence of the shock excitation of this PN (see also Lumsden et al. 2001).

The high R(H$_2$) values of the bipolar outflows of K~4-47 were unexpected. A comparison of our near-IR image with 
the optical images from Corradi et al. (2000) reveals that the ionized gas (optical emission) is concentrated in an 
inner highly collimated structure, which is surrounded by the H$_2$ bipolar outflows (Fig.~2). These structures of 
K~4-47 very closely resembles those of M~2-9 (Smith et al. 2005) and CRL~618 (Balick et al. 2013), suggesting a possible connection among these objects.

Recent hydrodynamic models by B. Balick and collaborators (Balick et al. 2013) provide a plausible explanation for the formation of the bipolar outflows in CRL~618. Naively comparing, the same model can adequately explain our findings for K~4-47. The H$_2$ emission at the walls of the bipolar outflows can be explained by the interactions between a jet or a bullet ejected from the central source with the surrounding AGB material. As the jet/bullet moves through the AGB material, it forms a conical structure that expands laterally outwards. At the same time, the jet/bullet continues moving outwards, with a velocity proportional to the distance from the central star, while dissociating the AGB H$_2$ gas, which then is ionized by the UV-photons emitted from the central star (optical emission). The models also predict strong near-IR \ironii\ $\lambda$1.644~$\mu$m emission at the outer layers of the knots, owing to the high expansion velocities. Unfortunately, the spectrum obtained by Lumsden et al. (2001) does not cover this line.

In view of the fact that the lateral expansion of the outflows and the velocity of the knots will slow 
down with time, we predict that the H$_2$ conical structure observed in K~4-47 will change with time to a more cylindrical shape,
similar to the H$_2$ structure in M~2-9 (Smith et al. 2005). This H$_2$ emission is, however, attributed to the stellar UV-photons (Kastner et al. 1996) rather than shock interactions. This difference from K~4-47 may be the result of the different evolutionary stage of the nebulae. 
In particular, K~4-47 has a kinematic age between 400 and 900~yr, depending on the distance (see Corradi et al. 2000), 
whereas M~2-9 is more evolved with a kinematic age of 2500~yr (e.g. Corradi et al. 2011). This may indicate that 
PNe like these two are predominantly shock-excited, and that as the central star evolves to a more luminous and hotter phase, 
the nebula becomes photoionized. By studying objects from different evolutionary stages, Gledhill and Forde (2015) reached the same conclusion. In particular, the Br$\gamma$ emission increases with time (e.g. IRAS 18062+2410), whereas the H$_2$ emission is stronger in the post-AGB phase 
and weaker in the more evolved phase of pre-PNe (e.g IRAS 20462+3416), or even absent in the most advanced phase of PNe (e.g. IRAS 19336-0400). 
This implies that the R(H$_2$) line ratio decreases with time. Theoretical models for different core masses have shown 
that the R(H$_2$) ratio is strongly dependent on time (evolutionary phase) (see Fig.~18 in Natta and Hollenbach 1998).

Regarding the elliptical PN NGC~7662, the H$_2$ emission was detected for the first time in LISs, whereas no emission associated with the nebular shells is found. This suggests that molecular H$_2$ gas in NGC~7662 is probably formed in clumpy structures, as LISs. 
Recently, van Hoof et al. (2010) investigated three different scenarios for H$_2$ formation in NGC~6720, through detailed photo-ionization 
modelling. These authors argued that the most plausible scenario for H$_2$ formation in NGC~6720 is inside the dense knots formed after the 
ionization phase, while they ruled out the scenario in which the H$_2$ gas survives in AGB knots. In contrast to this, it is curious to note that Matsuura et al. (2009) came to the exact opposite conclusion about the H$_2$ formation in the knots of the Helix nebula. If we compare the survival time of the knots of NGC~6720, 1500-15000~yr (as estimated by van Hoof et al. 2010), and the kinematical age of 
NGC~7662, $\sim$1600~yr (1050$\times \frac{D}{800}$, where D=1.2~kpc; Cahn et al. 1992), both scenarios are feasible for this nebula. 
Detailed modelling and deep spectroscopic data are necessary in order to obtain an accurate estimate of the survival times and the 
formation time-scales of the knots, which in turn will lead to a better understanding of how the H$_2$ condensations or LISs of NGC~7662 were formed.

\begin{figure}
\centering
\includegraphics[width=8.0 cm,height=7.75cm]{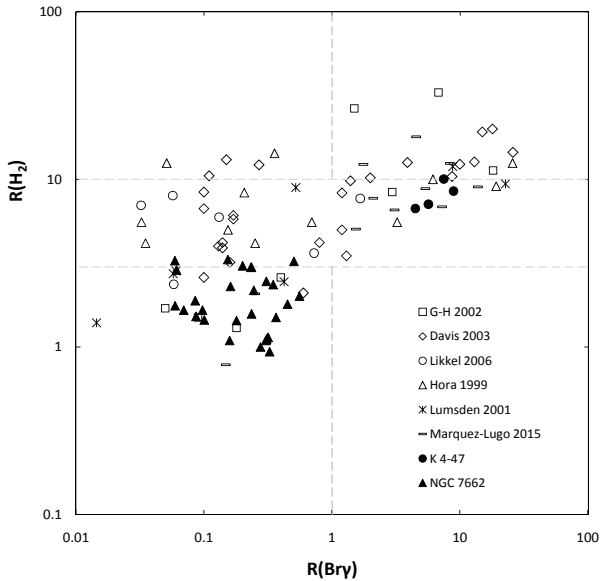}
\caption[]{R(H$_2$) versus R(Br$\gamma$) line ratio diagram from Marquez-Lugo et al. (2015) including the values of K~4-47 (filled circles) and NGC~7662 (filled triangles).}
\label{fig5}
\end{figure}

In contrast to the high-velocity knots in K~4-47, the LISs in NGC~7662 exhibit low expansion velocities of 30~\kms\ (Perinotto et al. 2004). 
Hence, the excitation of these LISs is probably the photoionization by the central star and not by shocks. 
This hypothesis is in agreement with the very low R(H$_2$) ratio of 2 to 3 found in this nebula (see Table~3).   
As noted above, R(H$_2$) alone cannot provide any robust excitation mechanism classification. 
Shock-excited regions are found to be systematically brighter in H$_2$ than in Br$\gamma$ (R(Br$\gamma$)$>$1), 
whereas  the photoionized regions are brighter in Br$\gamma$ (R(Br$\gamma$)$<$1) (Marquez-Lugo et al. 2015). This trend may reflect  
the correlation between the Br$\gamma$ line and the mass of the central star. In particular, massive stars or low-mass stars 
within the time interval from 1000 to 8000~yrs have R(Br$\gamma$)$<$1 (see Natta \& Hollenbach 1998). The low mass (0.605M$_\odot$) central star of NGC~7662 (Guerrero et al. 2004) and its kinematical age of $\sim$1600~yr imply a R(Br$\gamma$) lower than 1 
(see Fig. 18 in Natta \& Hollenbach 1998), in agreement with our results in Table~3.

In Fig.~4, we plot the R(H$_2$) versus R(Br$\gamma$) ratios of all the LISs in this work (K~4-47 and NGC~7662), together with the results from 
the census by Marquez-Lugo et al. (2015). We find that LISs in NGC 7662 are located in the locus of Br$\gamma$-bright source, 
where the photoionized regions are placed, whereas the bipolar outflows as well as the pair of fast-moving knots in K~4-47 are found in the regime 
of H$_2$-bright sources, where the main excitation mechanism is shocks. 

\begin{figure}
\centering
\includegraphics[width=8.5cm,height=7.0cm]{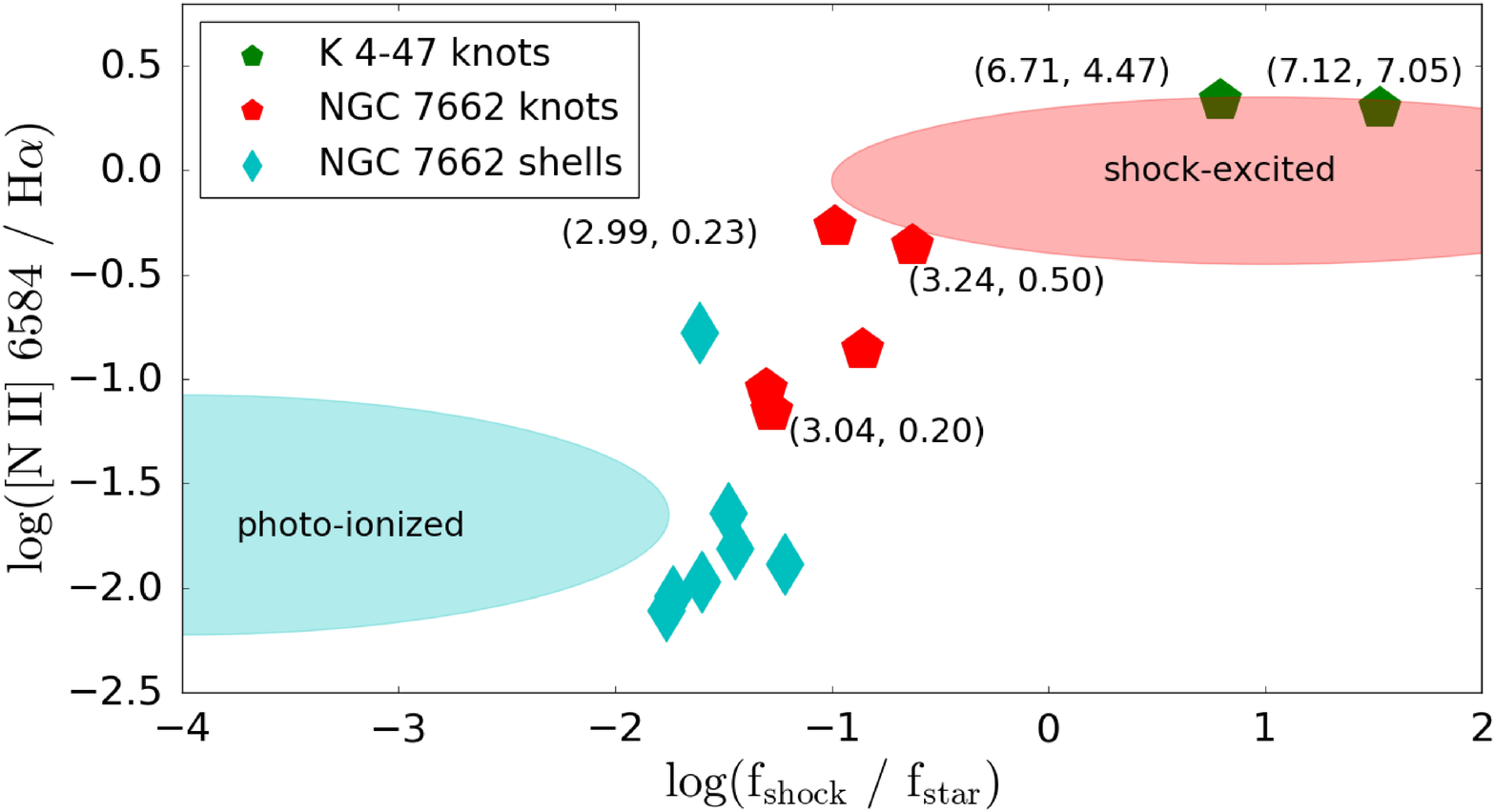}
\caption[]{ The diagnostic diagram for photo-ionized and shock-excited regions by Akras \& Gon\c{c}alves (2016). The data for K~4-47 and NGC~7662 were obtained from Gon\c{c}alves et al. (2009) and Perinotto et al. (2004), respectively. The numbers in parenthesis indicate the value of R(H$_2$) and R(Br$\gamma$). (A color version of this figure is available in the online journal.)} 
\label{fig5}
\end{figure}

In addition to the R(H$_2$) versus R(Br$\gamma$) plot, the new optical diagnostic diagram \nitrogen/H$\alpha$ versus 
log(f$_{\rm{shocks}}$/f$_{\rm{\star}}$) by Akras \& Gon\c{c}alves (2016) -- f$_{\rm{shocks}}$ and f$_{\star}$ are the ionization photon fluxes 
owing to the shocks and the central star ionizing continuum -- was also used to determine the excitation mechanism 
of LISs in K~4-47 and NGC~7662. The spectroscopic line ratios of the six LISs in this work (two knots of K~4-47 and 
four knots of NGC 7662) were gathered from the literature (Perinotto et al. 2004; Gon\c{c}alves et al. 2004, 2009). 
D=5.9~kpc and D=1.2~kpc are adopted for K~4-47 (Goncalves et al. 2004) and NGC 7662 (Cahn et al. 1992; Henry et al. 2015), respectively. 
We conclude that the knots of K~4-47 are predominantly shock-excited with log(f$_{\rm{shocks}}$/f$_{\rm{\star}})>$0, 
whereas  the LISs in NGC~7662 are found to lie in the transition zone with -2$<$log(f$_{\rm{shocks}}$/f$_{\rm{\star}})<$-1, 
where the two mechanisms, shocks and UV-photons, are equally important (Fig.~5). On other hand, regions of the outer shell and rim of NGC~7662 the cyan boxes in Fig.~5) are found to be very close to the regime of photoionized structures. This result strengthens our previous analysis that shocks have altered the ionization structure of LISs.

\section{Conclusion}

The deepest ever high-angular-resolution, near-IR narrow-band H$_2$ v=1-0 S(1), H$_2$ v=1-2 S(1) and Br$\gamma$ images of two Galactic PNe 
with LISs (K~4-47 and NGC~7662) have been presented in this work. H$_2$ was detected in both high-velocity knots (K~4-47) and low-velocity 
knots (NGC~7662) as well at the walls of the bipolar outflows of K~4-47. Therefore, we confirm that LISs are indeed partially made of H$_2$ gas, which,  in conjunction with their strong optical low-ionization emission lines, strongly suggest that LISs are probably mini-PDRs embedded in the nebulae. 
The H$_2$ emission was found to be excited either by shocks (K~4-47) or by high-energy 
UV-photons (NGC~7662). Moreover, the bright H$_2$ emission lines that emanated from the walls of the bipolar outflows of K~4-47 are consistent 
with the prediction of hydrodynamic models. In summary, these predictions are that a jet/bullet ejected from the central star interacts with the AGB remnant matter producing a conical structure that expands laterally. As the jet/bullet continues moving outwards, 
it dissociates the AGB materials, which later it is ionized by the UV radiation field of the central star. Moreover, we concluded that 
K~4-47 and M~2-9 may be akin young PNe in different evolutionary stage.

Additional deep (8~m class telescopes and similar), high-angular-resolution observations of PNe with LISs are required in order to cover a wider range of parameters (T$_{\rm{eff}}$, L, expansion velocities, n$_{\rm{e}}$) and provide a definitive understating of the excitation and formation mechanisms of the low-ionization structures of PNe.

\section*{Acknowledgments} 

We are extremely grateful to the anonymous referee for his constructive and helpful report. S.A. is supported by the CAPES Brazilian post-doctoral fellowship \lq\lq Young Talents Attraction" - Science Without Borders, A035/2013.  
GRL acknowledges support from CONACyT and PROMEP (Mexico).  
The observations we report here were obtained at the Gemini Observatory (processed using the Gemini IRAF package and Gemini-python), 
which is operated by the Association of Universities for Research in Astronomy, Inc., under a cooperative agreement 
with the NSF on behalf of the Gemini partnership: the National Science Foundation (United States), the National Research 
Council (Canada), CONICYT (Chile), the Australian Research Council (Australia), Minist\'{e}rio da Ci\^{e}ncia, 
Tecnologia e Inova\c{c}\~{a}o (Brazil) and Ministerio de Ciencia, Tecnolog\'{i}a e Innovaci\'{o}n Productiva (Argentina). 
This research is also based on observations made with the NASA/ESA Hubble Space Telescope, and obtained from the Hubble Legacy Archive, 
which is a collaboration between the Space Telescope Science Institute (STScI/NASA), the Space 
Telescope European Coordinating Facility (ST-ECF/ESA) and the Canadian Astronomy Data Centre (CADC/NRC/CSA). 

\bibliographystyle{mnras}

\newpage

\end{document}